\documentclass[prl,twocolumn,showpacs,floatfix]{revtex4-1}
\usepackage{amsmath}
\usepackage{amssymb}
\pdfoutput=1
\usepackage{graphicx}
\usepackage[english]{babel}
\usepackage{latexsym}
\usepackage{graphics}
\usepackage{subfigure}
\begin{document}
\title{Collective modes in two dimensional topological superconductors}
\author{Li Mao}
\thanks{Email: maoli@whu.edu.cn}
\author{Hongxing Xu}
\begin{abstract}
Collective modes in two dimensional topological superconductors are studied by an extended random phase approximation theory while considering the influence of vector field of light. In two situations, the s-wave superconductors without spin-orbit-coupling (SOC), and the hybrid semiconductor s-wave superconductor layers with strong SOC, we get the analytical results for longitudinal modes which are found to be indeed gapless. Further more, the effective modes volumes can be calculated, the electric and magnetic fields can be expressed as the creation and annihilation operators of such modes. So, one can study the interaction of them with other quasi-particles through fields. 
\end{abstract}
\affiliation{Department of Physics and Technology, Wuhan University, 
Wuhan, Hubei, 430072 CN}
\maketitle
\emph{Introduction} -- Recently, topological superconductor (TSC) has attracted great interests in condensed matter physics because of its rich physics \cite{Kitaev2001Unpaired,Read2000Paired,Alicea2011Non} and bright application prospects \cite{Kitaev2003Fault}. Similar to the topological insulator \cite{Qi2011Topological}, the quasi particle band structure of TSC exhibits none trivial topological structures. For example, in two dimensional (2D) TSC, the occupied states have odd Chern numbers ($\pm 1$ for p-wave or effective p-wave superconductors \cite{Read2000Paired} and $\pm 3$ for effective f-wave superconductors \cite{Mao2011Superconducting}). While in one dimensional (1D) systems, the quasi-particle band structures have $Z_2$ topological order \cite{Kitaev2001Unpaired}. The most interest and exotic property of TSC is that, in real space, there are Majorana zero energy modes (or Majorana fermions as quasi particle excitations of the system \cite{Majorana1937Teoria}) localized in the topology defects such as the vortex cores or edges in 2D systems. Zero energy means such excited states energetically degenerating to ground state, and the degenerated basis is proved to have non-Abelian statistics \cite{Alicea2011Non}. Which is supposed to be an excellent platform for fault-tolerant quantum computing \cite{Kitaev2003Fault}. 

Many condensed matter physics systems have been proposed to have TSC phases. Such as p-wave superconductors \cite{Ivanov2001Non}, hybrid structures consist of s-wave superconductor films and topological insulators \cite{Fu2008Superconducting} or semiconductor layers (n-type \cite{Sau2010Generic} and p-type \cite{Mao2011Superconducting}), and semiconductor wires placed on top of s-wave superconductors \cite{Lutchyn2010Majorana}. The Majorana zero energy modes have firstly been revealed by zero bias peaks in conductances of the N-NW-S (normal metal, semiconductor nanowire and superconductor) devices \cite{Mourik2012Signatures}. The fractional a.c. Josephson effect, which can be regarded as the responses of Majorana zero energy modes to external radiofrequency radiations, are then discovered \cite{Rokhinson2012}. We want to argue that, not only the states inside the superconducting gap (Majorana zero energy modes), but also the states outside it will respond to external fields. It is a collective oscillation of them. As we know, for an ordinary three dimensional (3D) s-wave superconductor, the energies of collective oscillations are too high to be influenced by the superconducting gap \cite{Anderson1958Random}. And for materials which can be regarded as layered superconductors, people have proved the collective excitation energy can be below the superconducting gap \cite{Fertig1990Collective}. So, will the collective modes should also be unaffected by the superconducting gap in 2D TSC is an important question when we investigate the responses to external fields.

Without superconductivity, collective modes in topological electron gases have been proposed to have lots of novel properties. For example, the so called Dirac plasmon, the collective mode of topological insulator surface states can be regarded as coupled electron and spin density oscillation due to the spin momentum locking effect \cite{Raghu2010Collective}. It has different dispersion relation compared to ordinary 2D electron gases, which has been proved by experiment \cite{DiPietro2013Observation}. It is well known that, when the time reversal symmetry (TRS) is broken, the TI surface states open a gap and have quantized Hall conductances \cite{Qi2008Topological}. At the same time, the surface states exhibit strong magnetoelectric effects \cite{Tse2010Giant}. None zero Hall conductance means the longitudinal and transverse responses can couple to each other, and makes the surface plasmon polaritons (pure TM modes without TRS breaking) acquiring a nonvanishing TE component \cite{Karch2011Surface}. So, what will happen when one add superconductivity to the system is another quite interesting question for collective modes in TSCs.

\begin{figure}[b]
\centering
\includegraphics[width=0.7\linewidth]{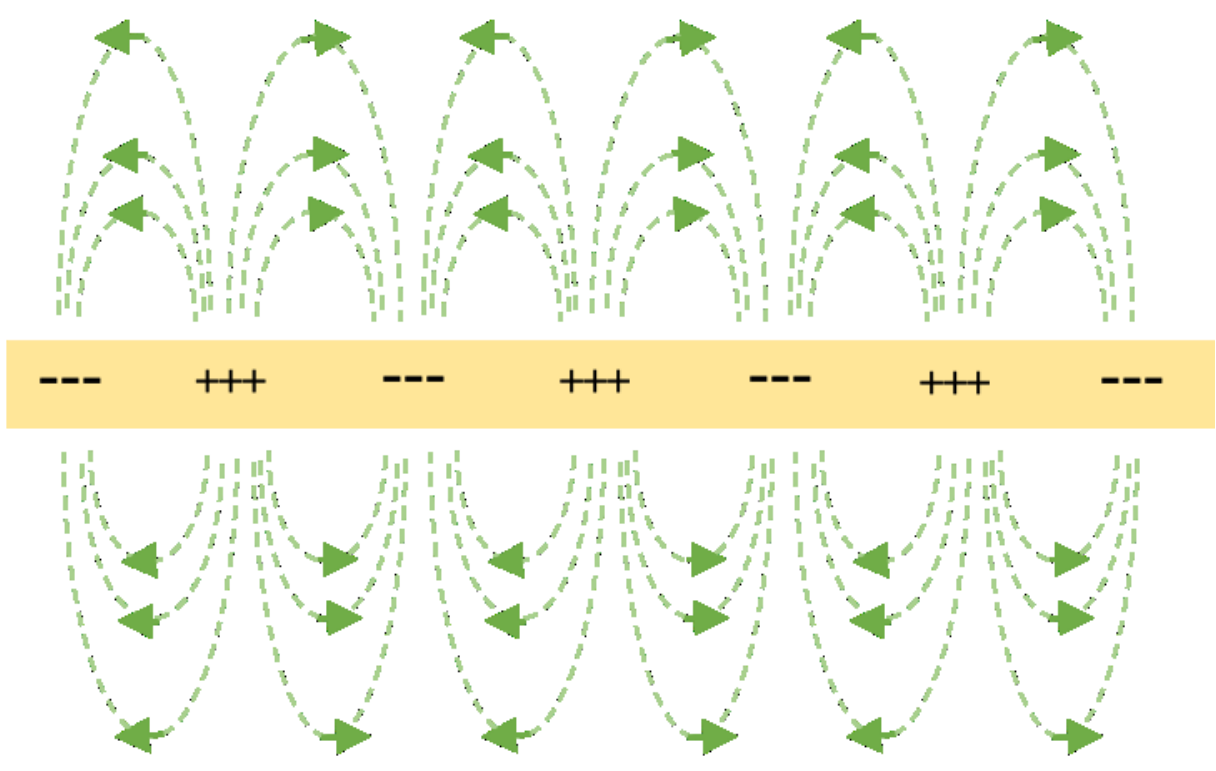} 
\caption{(Color online) Schematic picture for collective modes in a 2D topological superconductor.}
\end{figure}

Intrigued by these questions, we have studied the collective modes and none diagonal current current response function with an extended random phase approximation \cite{Anderson1958Random}. In order to study the interaction between collective modes and other quasi-particles, we calculate both the modes energies and wave functions with a method developed for none superconducting cases \cite{Mao2018Quantization}. This paper is organized as follows, firstly we express the collective modes as the summations of electron oscillations, photon annihilations and creations. By linearizing the dynamics of them, we can get the modes energies and wave functions. Then we calculate the interaction vertexes and the response functions for two special cases, and indeed find the collective modes are gapless. Finally we discuss their behaviors under topological phase transition. 

\emph{The dynamics of collective modes} --
The system we have considered can be expressed as Figure 1, where a 2D electron gas with strong SOC is putted inside a big box, the electrons interact to each other with Coulomb interaction and also through the vector field $\hat{A}(\mathbf{r})/c=1/\sqrt{\nu}\sum_{\mathbf{p}}e^{i\mathbf{p}\cdot\mathbf{x}}\sqrt{2\pi\hbar/\omega_{\mathbf{p}}}[\mathbf{\xi}_{\lambda}(\mathbf{p})\hat{a}^{\lambda}_{\mathbf{p}}(t)+\mathbf{\xi}_{\lambda}(-\mathbf{p})\hat{a}^{\lambda\dag}_{-\mathbf{p}}(t)]$, where we have used Einstein summation convention, $\hat{a}^{\lambda\dag}_{\mathbf{p}}$ and $\hat{a}^{\lambda}_{\mathbf{p}}$ are the free photon creation and annihilation operators in the box. $\mathbf{\xi}_{\lambda}(\mathbf{p})$ is the polarization vectors with 3D momentum $\mathbf{p}$. The total Hamiltonian reads $\hat{\mathbf{H}}(t)=\hat{\mathbf{H}}_{BdG}(t)+\hat{\mathbf{H}}_j(t)+\hat{\mathbf{H}}_{ee}(t)-\hat{\Sigma}(t)$. 
\begin{align}
&\hat{\mathbf{H}}_{BdG}(t)=\int d\mathbf{x}
\hat{\Psi}^{\dag}(\mathbf{x},t)
\begin{pmatrix}
H_s(\mathbf{\hat{p}})
& \Delta\sigma_0\\ \Delta\sigma_0 &
-\sigma_2 H^{*}_s(\mathbf{\hat{p}}
)\sigma_2
\end{pmatrix}
\hat{\Psi}(\mathbf{x},t)\nonumber\\
&\hat{\mathbf{H}}_{j}(t)=-\frac{1}{c}\int 
[\hat{\mathbf{j}}^{p}(\mathbf{x},t)
+\frac{1}{2}\hat{\mathbf{j}}^{d}(\mathbf{x},t)]\cdot
\hat{\mathbf{A}}(\mathbf{x},t)d\mathbf{x},\nonumber\\
&\hat{\mathbf{H}}_{ee}(t)=\frac{1}{2}\int d\mathbf{x}d\mathbf{x}'
\hat{\Psi}^{\dag}(\mathbf{x},t)\kappa_{30}
\hat{\Psi}(\mathbf{x},t)[V_c(|\mathbf{x}-\mathbf{x}'|)\nonumber\\
&\quad\quad\quad-g\delta(\mathbf{x}-\mathbf{x}')]\hat{\Psi}^{\dag}(\mathbf{x}',t)\kappa_{30}
\hat{\Psi}(\mathbf{x}',t),\nonumber\\
&\hat{\Sigma}(t)=\int d\mathbf{x}
\hat{\Psi}^{\dag}(\mathbf{x},t)\kappa_{10}\Delta
\hat{\Psi}(\mathbf{x},t)
\end{align}
where we have defined $\kappa_{ij}=\sigma_i\otimes\sigma_j$, $\hat{\Psi}(\mathbf{x},t)=1/\sqrt{2}[\hat{\Psi}_{\uparrow}(\mathbf{x},t),\hat{\Psi}_{\downarrow}(\mathbf{x},t),\hat{\Psi}^{\dag}_{\downarrow}(\mathbf{x},t),-\hat{\Psi}_{\uparrow}^{\dag}(\mathbf{x},t)]^{T}$ is the Nambu spinor \cite{Nambu1960Quasi}, $H_s(\hat{\mathbf{p}})=(\hat{\mathbf{p}}^2/2m^*-\mu)\sigma_0+\alpha (\hat{\mathbf{p}}_x\sigma_2-\hat{\mathbf{p}}_y\sigma_1)+h\sigma_3$ is the single electron Hamiltonian, $\hat{\mathbf{j}}^{p}(\mathbf{x},t)$ and $\hat{\mathbf{j}}^{d}(\mathbf{x},t)$ are paramagnetic and diamagnetic currents (The detail formulas can be found in the supplementary materials). $-g\delta(\mathbf{x}-\mathbf{x}')$ is the s-wave attractive interaction induced by exchanging phonon or other mechanisms. $\hat{\mathbf{H}}_{BdG}(t)$ is the Bogoliubov de Gennes (BdG) Hamiltonian for superconductors, $\hat{\mathbf{H}}_j(t)$ is the Hamiltonian of vector field coupled to electric current, and $\hat{\mathbf{H}}_{ee}(t)$ is the effective electron electron interaction term (contains the Coulomb interaction and an attractive interaction induced by exchanging phonons). Because the none diagonal self energy $\hat{\Sigma}(t)$ has been added in the BdG Hamiltonian, so we subtract it in the interaction terms.

Then the collective modes can be expressed as
\begin{align}
\hat{Q}^{\dag}_{\mathbf{q}}=&\sum_{\mathbf{k}}\hat{\Psi}^{\dag}_{\mathbf{k}+\mathbf{q}}\Phi^{\rho}\hat{\Psi}_{\mathbf{k}}+\sum_{p_z}\Phi^a_{\lambda}\hat{a}^{\lambda}_{-\mathbf{q},-p_z}+\Phi^c_{\lambda}\hat{a}^{\lambda\dag}_{\mathbf{q},p_z}
\end{align}
\begin{figure}[t]
  \centering
  \includegraphics[width=8cm,bb=0 0 600 140]{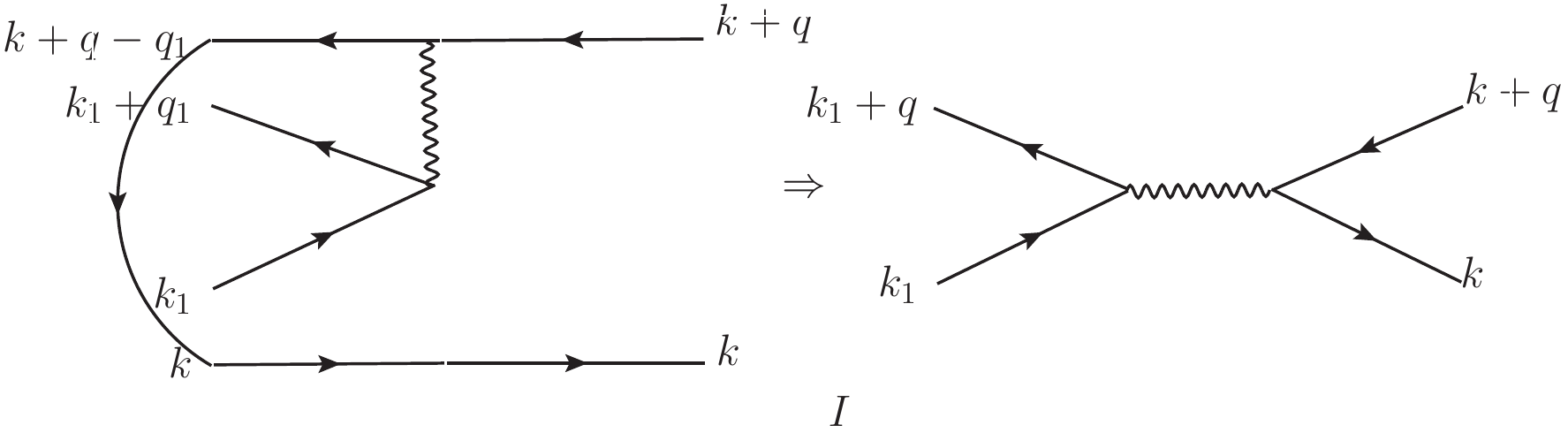}
  \includegraphics[width=8cm,bb=0 0 600 160]{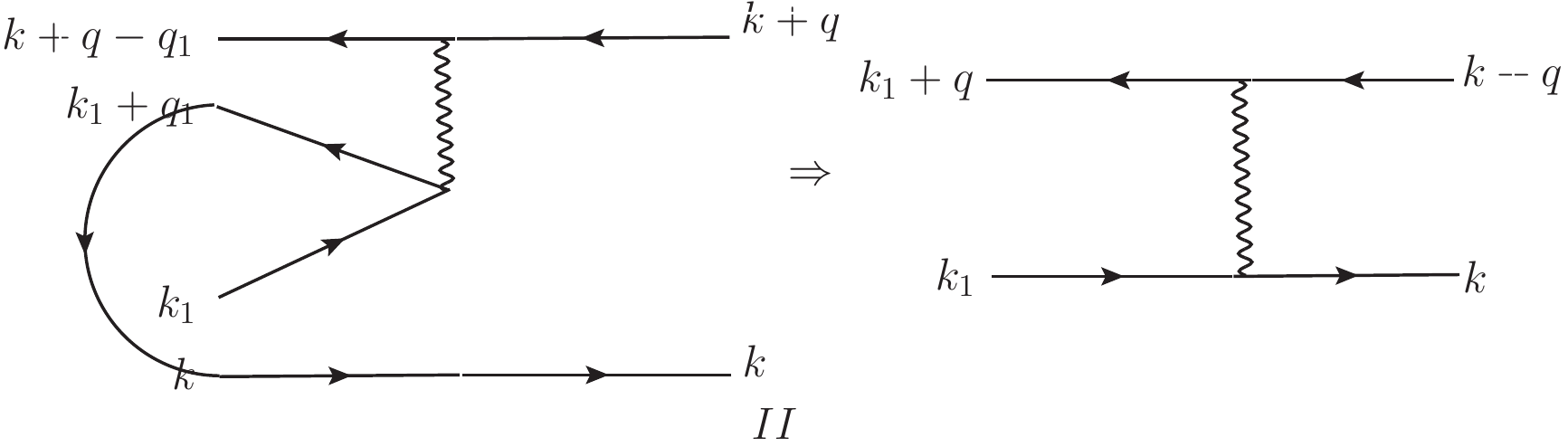}
 \caption{Contributions of $\hat{\mathbf{H}}_{ee}(t)$ to the quasi particle's dynamics. I, the ordinary random phase approximation, II, the ladder approximation.}
  \label{fig:2}
\end{figure}

Now $\mathbf{k},\mathbf{q}$ are in-plane momentum, $p_z$ is the momentum perpendicular to them. We can apply an extended random phase approximation, replacing quasi particle hole pairs with their ground state average values, to linearize the dynamic equation $i\hbar\partial_t\hat{Q}^{\dag}_{\mathbf{q}}=[\hat{Q}^{\dag}_{\mathbf{q}}, \hat{\mathbf{H}}(t)]=-\hbar\hat{Q}^{\dag}_{\mathbf{q}}$. For the electron part, the contributions of $\hat{\mathbf{H}}_{ee}(t)$ can be illustrated by figure 2 and 3 (connecting the particle and hole lines means taking their average values). In figure 2, I is just the ordinary random phase approximation (the bubble terms), II is originated by the attractive interaction. In figure 3, III and IV can be regarded as self energy terms and exactly cancel the contribution of $\hat{\Sigma}(t)$. After lots of calculations (please check the supplementary materials), we find the wave function can be expressed as  
\begin{figure}
  \centering
  \includegraphics[width=8cm,bb=0 0 600 140]{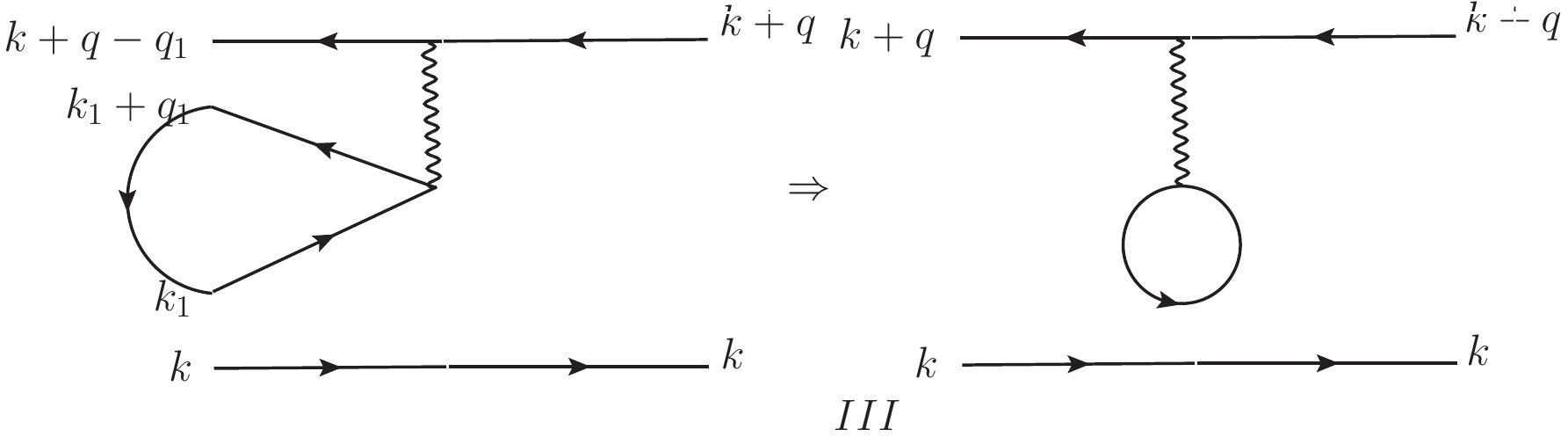}
  \includegraphics[width=8cm,bb=0 0 600 160]{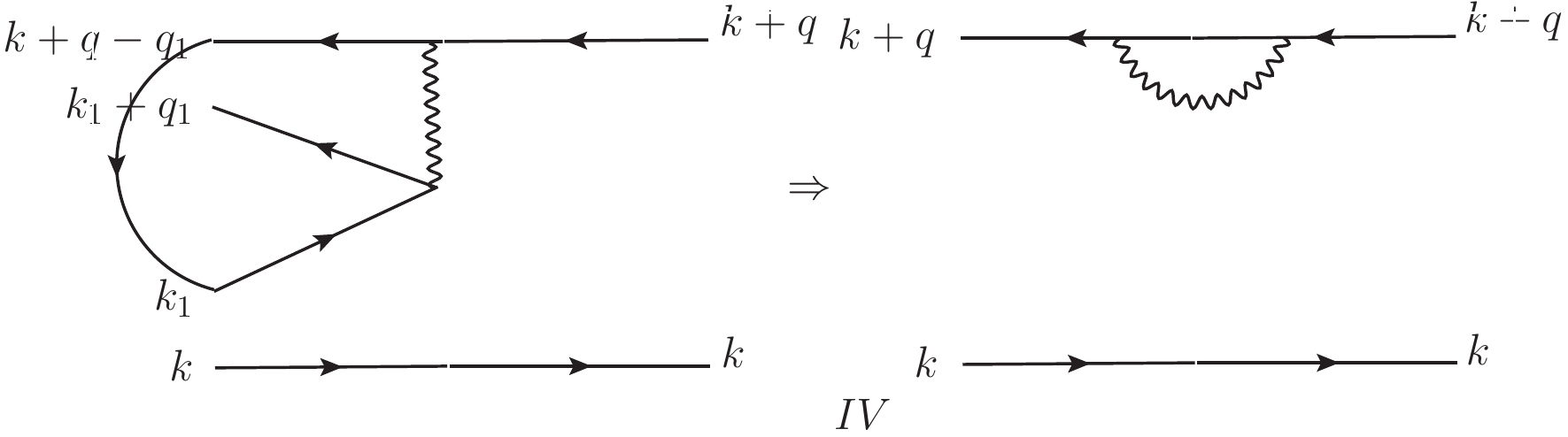}
 \caption{Contributions of $\hat{\mathbf{H}}_{ee}(t)$ to the quasi particle's dynamics. III and IV, the 'self energy' terms.}
  \label{fig:3}
\end{figure}
\begin{align}
\Phi^{\rho}
=\frac{|s\mathbf{k}+\mathbf{q}\rangle\langle s\mathbf{k}+\mathbf{q}|\Gamma_iL^{i}|s'\mathbf{k}\rangle\langle s'\mathbf{k}|}{\hbar\Omega_{\mathbf{q}}-\Delta\xi^{ss'}_{\mathbf{k}\mathbf{q}}+i\delta}
\end{align} 
 where $\Gamma_i$ ($i=\rho,\ j_x,\ j_y$) is the dressed vertex, their relations to the bare vertex, $\gamma_i$, reads
 \begin{align}
&\Gamma_{i}=\gamma_i+\frac{1}{\beta\hbar S}\sum_{i\omega_n,\mathbf{k}}
[\kappa_{30} G(i\omega_n+i\Omega_{\mathbf{q}},\mathbf{k}+\mathbf{q})
\Gamma_{i}\nonumber\\
&\quad\quad\ \times g(\mathbf{k},\mathbf{q})G(i\omega_n,\mathbf{k})\kappa_{30}]_{i\Omega_{\mathbf{q}}\rightarrow \Omega_{\mathbf{q}}}\nonumber\\
&\gamma_{\rho}=\kappa_{30},\ 
\gamma_{j_x}=\frac{\hbar}{m^*}(k_x+q/2)\kappa_{00}-\hbar\alpha \kappa_{02},
\nonumber\\ 
&\gamma_{j_y}=\frac{\hbar}{m^*}k_y\kappa_{00}-\hbar\alpha \kappa_{02},
\end{align}
And $L^i$ are normalization factors that satisfy the following linear equations
\begin{align}
&L^{i}=V_i(q)\Pi_{ij}(\Omega_{\mathbf{q}},\mathbf{q})L^{j},
 \end{align}
 $V_{\rho}(q)=2\pi e^2/q$ is the static photon propagator, $V_{j_x}(q)=V_{\mathbf{q}}^l=2\pi e^2q'\omega^2$ and $V_{j_y}(q)=V_{\mathbf{q}}^t=-2\pi e^2/(c^2q')$ are longitudinal and transverse photon propagators while $q'=\sqrt{q^2-\Omega^2_q/c^2}$. 
 Immediately, one can find the frequency should be determined by
 \begin{align}
V_{q}^lV_{q}^t|\Pi_{j_x j_y}|^2=(1-V_{q}^l\Pi_{j_xj_x})(1-V_{q}^t\Pi_{j_yj_y})
 \end{align}
 \emph{Collective modes in s-wave superconductors} -- 
 As a simple example, we calculate the collective modes for s-wave superconductors without SOC and Zeeman field. Now, the Nambu spinor reduced to $\hat{\Psi}=[\hat{\Psi}_{\mathbf{k}\uparrow},\hat{\Psi}^{\dag}_{-\mathbf{k}\downarrow}]$. The single particle Hamiltonian in $\mathbf{k}$ space reads $H_s(\mathbf{k})=(\hbar k^2/2m^*-\mu)\sigma_0$, so the free vertexes $\gamma_{j_x}=\hbar(k_x+q/2)/m^*\sigma_{0}$ and $\gamma_{j_y}=\hbar k_y/m^*\sigma_0$. The Green function reads
\begin{align}
G_{nsoc}(i\omega_n,\mathbf{k})=\sum_{s=\pm 1}
  \frac{1}{2E_k}\frac{E_k\sigma_0+s\alpha\xi_k\sigma_3+s\Delta \sigma_1}{i\omega_n-sE_k}
\end{align}
By expanding the vertex $\Gamma_i=\sigma_l \Gamma_i^{l}$, and defining $\tilde{\Gamma}_i=[\Gamma_i^{0},\Gamma_i^{1},\Gamma_i^{2},\Gamma_i^{3}]^T$ and $\tilde{\gamma}_i$ in the same way, we find 
\begin{align}
\tilde{\Gamma}_i=\tilde{\gamma}_i+M\tilde{\Gamma}_i
\end{align}
 with $M^{lm}=1/2S\sum_{i\omega_n,\mathbf{k}}Tr[\sigma_l\sigma_3 G_{nsoc}(i\omega_n+i\Omega_{\mathbf{q}},\mathbf{k}+\mathbf{q})\sigma_mG_{nsoc}(i\omega_n,\mathbf{k})\sigma_3]$. 

 One can easily check $\Pi_{j_xj_y}=0$, which means the longitudinal and transverse modes are not coupled to each other. In the limit $\mu>>\Delta$ and $q\rightarrow 0$, $\theta_{\mathbf{k}+\mathbf{q}}=\theta_{\mathbf{k}}+O(q/k_f)$. The density-density response function now reads 
 \begin{align}
\Pi_{\rho\rho}
\approx&
2\frac{[1+gF_{++}(\mathbf{q})]F_{-+}(\mathbf{q})-gA^2_{++}(\mathbf{q})}
{g^2A^2_{++}(\mathbf{q})+(1-gF_{-+}(\mathbf{q}))(1+gF_{++}(\mathbf{q}))}
\end{align}
where
\begin{align}
&\ A_{ss'}(\mathbf{q})=\frac{1}{4\pi^2}\int A_{s}(\mathbf{k},\mathbf{q})L_{s'}(\mathbf{k},\mathbf{q})d\mathbf{k},\nonumber\\
&F_{ss'}(\mathbf{q})=\frac{1}{4\pi^2}\int F_{ss'}(\mathbf{k},\mathbf{q})L_-(\mathbf{k},\mathbf{q})d\mathbf{k},\nonumber\\
&A_{s}=\frac{E_{k+}+sE_{k-}}{4E_{k+}E_{k-}}\Delta,\nonumber\\
&B_s=\frac{\xi_{k+}+s\xi_{k-}}{4E_{k+}E_{k-}}\Delta\nonumber\\
&C_s=\frac{E_{k+}\xi_{\mathbf{k}-}+sE_{k-}\xi_{\mathbf{k}+}
}{4E_{k+}E_{k-}}\nonumber\\
&F_{ss'}=\frac{E_{k+}E_{k-}+s\xi_{k+}\xi_{k-}+s'\Delta^2}{4E_{k+}E_{k-}},\nonumber\\
&L_s=\frac{1}{\omega_q-E_{k+}-E_{k-}}+s\frac{1}{\omega_q+E_{k+}+E_{k-}}.
\end{align}
If we ignore the retardation effect $q'\approx q$, we can find the longitudinal collective modes are determined by $1-V_q\Pi_{\rho\rho}=0$, and finally get $\Omega_q=\sqrt{2\mu e^2q}=\sqrt{2\pi e^2n/m^*}\sqrt{q}$. Which means the plasmon frequency is unaffected by the superconducting gap to the order of $O(q^{1/2})$.

\emph{Collective modes in TSCs} -- 
Then let's calculate the collective modes for the semiconductor/s-wave superconductor hybrid system with strong SOC. One can find when the Zeeman filed $h$ is large enough, the bands are well separated at $k=0$, while the band gap around $k_f=2m^*\alpha/\hbar$ keep unchanged if $\alpha$ is big enough. Or, when $\hbar\alpha k_f>>h>>\Delta$, and $\hbar\alpha k_f>>\omega_D$, only the integrating near $k_f$ contribute to the collective modes and the self consistent equation of $\Delta$. So, we can develop a much simpler theorem by expanding Hamiltonian near $k_f$. Now, the Green function reads
\begin{align}
  G(i\omega_n,\mathbf{k})=&\frac{1}{2}G_{nsoc}(i\omega_n,\mathbf{k})\otimes (\sigma_0+\sigma_1\sin{\theta_{\mathbf{k}}}\nonumber\\
  &-\sigma_2\cos{\theta_{\mathbf{k}}})
\end{align}
Note that, in $G_{nsoc}(i\omega_n,\mathbf{k})$, we replace $\xi_{\mathbf{k}}$ with $\hbar \alpha \delta k=\hbar \alpha(k-k_f)$.
Similar to the none SOC case, expand the vertex $\Gamma_i=\kappa_{nm}\Gamma_i^{nm}$. Immediately, one can find for $m=3$, $\Gamma_i^{n3}=0$. So, let's define $\tilde{\Gamma}_i=[\Gamma_i^{00},\Gamma_i^{01}\cdot\cdot\cdot\Gamma_i^{32}]^T$ and $\tilde{\gamma}_i$ in the same way. Then we find 
\begin{align}
\tilde{\Gamma}_i=\tilde{\gamma}_i+M\otimes S\tilde{\Gamma}_i
\end{align}
where (to the second order of $q$)
\begin{align}
&M(q)=\left(\begin{array}{cccc}
F_{--}(\mathbf{q})
& -A_{-+}(\mathbf{q})
&iB_{--}(\mathbf{q})
&-C_{-+}(\mathbf{q})\\
A_{-+}(\mathbf{q})
&-F_{+-}(\mathbf{q})
&iC_{++}(\mathbf{q})
&B_{+-}(\mathbf{q})\\
iB_{--}(\mathbf{q})
&-iC_{++}(\mathbf{q})
&-F_{++}(\mathbf{q})
&iA_{++}(\mathbf{q})\\
-C_{-+}(\mathbf{q})
&-B_{+-}(\mathbf{q})
&iA_{++}(\mathbf{q})
&F_{-+}(\mathbf{q})
\end{array} \right)\nonumber\\
&S=\begin{pmatrix}
    1 & \sin{\theta_k} &-\cos{\theta_k}\\
   \sin{\theta_k} & \sin^2{\theta_k} &-\sin{\theta_k}\cos{\theta_k}\\
    -\cos{\theta_k} & -\sin{\theta_k}\cos{\theta_k} &\cos^2{\theta_k}
\end{pmatrix}
\end{align}
And one can find
\begin{align}
\Gamma_{\rho}=&-\frac{gk_fq}{4\pi\Omega_q^2}\kappa_{02}
+\frac{g\alpha^2k_f^2q^2}{4\pi\Delta\Omega_q^2}\kappa_{10}
+i(\frac{2\Delta}{\Omega_q}-\frac{g\Delta^2k_fq}{2\pi\Omega_q^3})\kappa_{20}\nonumber\\
&+(1-\frac{g\alpha k_fq^2}{4\pi\Omega_q^2})\kappa_{30}
\end{align}
and find the density density response function $\Pi_{\rho\rho}=\alpha k_fq^2/(2\pi\Omega_q^2)$. So, by solving $1-V(q)\Pi_{\rho\rho}=0$, we can get the collective mode energy $\Omega_q=\alpha e/\hbar\sqrt{2m^*/\epsilon_0q}$. Note that it is proportional to $\sqrt{q}$, which means our expanding method is self consistent. Finally, about the none diagonal current current response function, we find it is zero in this case. Note that we have ignored $O(1/k_f)$ terms, so we need to consider higher order terms in order to investigate modes hybridization under topological phase transition. 

\emph{Fields of collective modes} -- Once getting the modes energy, we can calculate the normalization factors by the requirement that the collective modes should be bosons $[\hat{Q}_{\mathbf{q}},\ \hat{Q}_{\mathbf{q}'}]=\delta_{\mathbf{q}\mathbf{q}'}$. Then we can get the inverse transformation which represents electron oscillations and photons with collective modes \cite{Mao2018Quantization}. For the photon parts, we have
$\hat{a}^{\lambda}_{-\mathbf{q},-p_z}=-\Phi_{\lambda,p_z}^{a *}
\hat{Q}^{\dag}_{\mathbf{q}}+\Phi^{b}_{\lambda,-p_z}\hat{Q}_{-\mathbf{q}}$ and
$\hat{a}^{\dag}_{\lambda\mathbf{q}p_z}=\Phi_{\lambda,p_z}^{b *}
\hat{Q}^{\dag}_{\mathbf{q}}-\Phi^{a}_{\lambda,-p_z}\hat{Q}_{-\mathbf{q}}$.
Which means the vector fields have the following form,
\begin{align}
\hat{A}_{z}(\mathbf{r},z)=&\frac{ic}{e}\sum_{\mathbf{q}}e^{-i\mathbf{q}\cdot\mathbf{r}}
\frac{L^{\rho *}_{\mathbf{q}}q}{\omega_{\mathbf{q}}}(e^{-qz}-e^{-q'z})
\hat{Q}^{\dag}_{\mathbf{q}}+h.c.\nonumber\\
\hat{A}_{\parallel}(\mathbf{r},z)=&-\frac{c}{e}\sum_{\mathbf{q}}\frac{1}{\Omega_{\mathbf{q}}}e^{-i\mathbf{q}\cdot\mathbf{r}}
[L^{\rho *}_{\mathbf{q}}
(\mathbf{q}e^{-qz}
-\mathbf{q}'e^{-q'z})\nonumber\\
&+T^*_{\mathbf{q}}e^{-q'z}
\tilde{\mathbf{q}}']\hat{Q}^{\dag}_{\mathbf{q}}+h.c.
\end{align}
where $\tilde{\mathbf{q}}$ is an in-plane vector perpendicular to 
$\mathbf{q}$ ($\theta_{\tilde{\mathbf{q}}}=\theta_{\mathbf{q}}+\pi/2$), and $\mathbf{q}'$ means the magnitude of 
$\mathbf{q}$ reduced from $q$ to $q'=\sqrt{q^2-\Omega^2_{\mathbf{q}}/c^2}$, 
and $T_{\mathbf{q}}$ is another wave function normalization factor 
existing in transverse or longitudinal transverse hybrid modes. 

Also, we can get the effective static Coulomb potential generated by collective mode $\hat{V}_{c}^{spp}=\sum_{\mathbf{q}}e^{-i\mathbf{q}\cdot\mathbf{r}-qz}L_{\mathbf{q}}^{*}\hat{Q}_{\mathbf{q}}^{\dag}+h.c.$. Then one can study the interaction between collective mode and other quasi-particle by electric field
\begin{align}
&\hat{E}(\mathbf{x})=-\frac{1}{c}\frac{\partial \hat{A}(\mathbf{x})}{\partial t}+\frac{1}{e}\nabla \hat{V}^{spp}_{c}
(\mathbf{x})\nonumber\\
&
=-\frac{1}{e}\sum_{\mathbf{q}}
e^{-i\mathbf{q}\cdot\mathbf{r}-q'z}[(i\mathbf{q}',q)L^{\rho *}_{\mathbf{q}}
+\Omega_{\mathbf{q}}\frac{\tilde{\mathbf{q}}}{q}
T^*_{\mathbf{q}}]\hat{Q}_{\mathbf{q}}^{\dag}+h.c..
\end{align}
\emph{Conclusion} -- The collective modes of 2D TSC have been studied with an extended random phase approximation, both the modes energy and wave function can be calculated by this method. So, one can express the electric field generated by them with quasi particles. For two special cases, the SOC is absent in the system and with strong Zeeman field and SOC, we get analytic results for the modes energies, and find them are indeed gapless. Which means they are quite important when one consider the responses of TSC to external fields. 

This work is supported by the Ministry of Science and Technology 
(Grant 2015CB932401), the National Key R\&D Program of China (Grant
2017YFA0205800, 2017YFA0303402), the National Natural Science Foundation of 
China (Grant 11344009).

\bibliography{cmtsc} 
\end{document}